\documentclass[12pt]{article}

\advance \textheight by \topskip
\makeatletter
\@addtoreset{equation}{section}
 
\makeatother

\def\N{\mathbb N}
\def\Z{\mathbb Z}

\usepackage{amsmath,amssymb}

\begin{document}

\title{
{\bf Nonlinear superconformal symmetry
of a fermion in the field of a Dirac monopole}}

\author{{\sf Carlos Leiva${}^{a}$}\thanks{
E-mail: caleiva@lauca.usach.cl}
{\sf\ and Mikhail S. Plyushchay${}^{a,b}$}\thanks{
E-mail: mplyushc@lauca.usach.cl}
\\
{\small {\it ${}^a$Departamento de F\'{\i}sica,
Universidad de Santiago de Chile,
Casilla 307, Santiago 2, Chile}}\\
{\small {\it ${}^b$Institute for High Energy Physics,
Protvino, Russia}}}
\date{}






\maketitle


\begin{abstract}
We study a longstanding problem of
identification of the fermion-monopole symmetries.
We show that the integrals of motion of the
system generate a nonlinear classical $\Z_2$-graded Poisson,
or quantum super- algebra, which may be treated as a
nonlinear generalization of the
$osp(2|2)\oplus su(2)$.
In the nonlinear superalgebra,
the shifted square of the full angular momentum
plays the role of the central charge.
Its square root is the  even $osp(2|2)$ spin generating
the $u(1)$ rotations of the supercharges.
Classically,
the central charge's square root
has an odd counterpart
whose quantum analog is, in fact,  the same
$osp(2|2)$ spin operator.
As  an odd integral, the
$osp(2|2)$ spin generates a nonlinear supersymmetry
of De Jonghe, Macfarlane, Peeters
and van Holten, and
may be identified as a grading operator of the
nonlinear superconformal algebra.
\end{abstract}

\newpage
\section{Introduction}
Hidden, or dynamical symmetries are behind
the special properties of some classical and quantum
mechanical systems. The best known example of
the hidden symmetry is provided, probably,
by the Kepler problem.
Being associated with the Laplace-Runge-Lenz vector
integral, it is  responsible for the closed  character of
the finite orbits,
fixing the orbits' orientation
in the both cases of a finite and an infinite motion.
Also, it explains a degeneracy of the quantum spectrum.

The problem of identification of
symmetries of the fermion-monopole system
is a long-standing puzzle.
More than twenty years ago, Jackiw
found that the system of a scalar charged particle
in the field of a Dirac magnetic monopole
possesses a hidden
conformal $so(1,2)$ symmetry \cite{Jackiw}.
Four years later D'Hoker and Vinet showed that the
fermion-monopole system can be characterized by the
$osp(1,2)$ supersymmetry \cite{DHV}.
Then, De Jonghe, Macfarlane, Peeters and van
Holten \cite{MacHol} identified a new supersymmetry of the
system which squares to the
(shifted) Casimir invariant of the full rotation group.
Their analysis,
based on a general method which uses
Killing-Yano tensors to generate
additional supersymmetries \cite{GRH},
revealed that
the full supersymmetry algebra of
the fermion-monopole is, in fact,
nonlinear.
Lately, Spector \cite{Spec} found  that such a non-standard
nonlinear supersymmetry  associated with the Casimir
invariant of the
full rotation group may exist
even when an additional potential term
breaking a usual supersymmetry
is included in the Pauli Hamiltonian.
In both papers \cite{MacHol,Spec}, the supersymmetries
were treated outside the context of conformal
symmetry. Following their line, in Ref. \cite{PlFM}
it was argued that the commutator of the usual
(``square root from the Hamiltonian")
and of the non-standard
(``square root from the rotational
Casimir") supercharges should also be treated
as a new supercharge, which squares for the
product of the Hamiltonian and rotational
Casimir
invariant\footnote{See Eq. (\ref{Kep})
and the comment to it.}.
This resembles
the nonlinear structure of
the Kepler problem, where
the commutator of the Laplace-Runge-Lenz
vector components takes a form of
the product of the Hamiltonian
and angular momentum operator
\cite{nonlin}.

The $osp(1,2)$ symmetry of the fermion-monopole system
\cite{DHV} supplied  with the additional nonlinear
supersymmetry of Refs. \cite{MacHol,PlFM}
is in the obvious but puzzling
contrast with the
$osp(2,2)$ supersymmetry of the
Akulov-Pashnev-Fubini-Rabinovici
superconformal model
\cite{AP,FR,IKL,scf}.
The difference is rather surprising
since the both systems
are characterized by the same
Lie algebraic dynamical
structure $so(1,2)$, and
their quantum spaces
associated with the spin
degrees of freedom are exactly the same.
Recently, however, it was observed
\cite{LP,AnPl}
that for the special (discrete)
values of the fermion-boson coupling
parameter,
in addition to the usual $osp(2,2)$ symmetry,
the superconformal model \cite{AP,FR}
can be  characterized by
a hidden nonlinear superconformal symmetry.

In the present paper, we  investigate
the problem of identification of
(super)symmetries of the fermion-monopole
system by comparing its structure
with realization of superconformal symmetry in the
model \cite{AP,FR}.
As a result, it will be shown that the system
possesses a nonlinear
superconformal symmetry.
A nonlocal transformation applied
to the states
with the angular momentum different from the lowest
one, reduces
the full symmetry to the usual $osp(2,2)$ supersymmetry
(plus a `decoupled' rotational symmetry) ,
in which the square root from the
shifted rotational Casimir
invariant plays the role of the central charge
similar to the boson-fermion coupling
constant of the superconformal model \cite{AP,FR}.

The paper is organized as follows.
In section 2 we analyze the dynamical symmetry of the
superconformal mechanics model. The results of the analysis
are applied in section 3 to find all the set of the
integrals generating the dynamical symmetry of the
fermion-monopole system and to identify the corresponding
superalgebra at the classical and quantum levels.
In section 4 we summarize the obtained results and discuss
some open problems to be interesting for further
investigation.


\section{Dynamical symmetry of the 1D superconformal
mechanics}
We start with a short discussion
of the symmetries of the 1D
superconformal model \cite{AP,FR,IKL,scf},
that will be helpful for subsequent identification
of the fermion-monopole symmetries.

The classical model \cite{AP,FR} is given by the Hamiltonian
\begin{equation}
H=\frac{1}{2}\left(p^2+
\frac{\alpha(\alpha+2i\psi_1\psi_2)}{x^2}\right),
\label{hscm}
\end{equation}
and by the fundamental Poisson brackets
$\{x,p\}=1$, $\{\psi_a,\psi_b\}=-i\delta_{a,b}$, $a,b=1,2$.
In addition to $H$,
the system  possesses the even,
\begin{equation}
D=
\frac{1}{2}xp-tH,\qquad
K=
\frac{1}{2}x^2-2tD-t^2H,\qquad
\Sigma=-i\psi_1\psi_2,
\label{tdk}
\end{equation}
and odd,
\begin{equation}
Q_a=p\psi_a+\frac{\alpha}{x}\epsilon_{ab}\psi_b,\qquad
\tilde{Q}_a=x\psi_a-tQ_a,
\label{qs}
\end{equation}
integrals of motion obeying the equation
of the form
$\frac{d}{dt}I=\frac{\partial}{\partial t}I+\{I,H\}=0$.
The set of integrals (\ref{hscm}), (\ref{tdk}),
(\ref{qs}) generates the $osp(2,2)$
superalgebra given by the nontrivial Poisson bracket
relations
\begin{eqnarray}
&\{H,K\}=-2D,\quad
\{D,H\}=H,\quad
\{D,K\}=-K,&\nonumber
\\
&\{Q_a,Q_b\}=-2i\delta_{ab}H,\quad
\{\tilde{Q}_a,\tilde{Q}_b\}=-2i\delta_{ab}K,&\nonumber
\\
&\{\tilde{Q}_a,Q_b\}=-2i\delta_{ab}D+i\epsilon_{ab}(\Sigma+
\alpha),&
\label{osp22}
\\
&\{H,\tilde{Q}_a\}=-Q_a,\quad
\{K,Q_a\}=\tilde{Q}_a,\quad
\{D,Q_a\}=\frac{1}{2}Q_a,\quad
\{D,\tilde{Q}_a\}=-\frac{1}{2}\tilde{Q}_a,&\nonumber
\\
&\{\Sigma,Q_a\}=\epsilon_{ab}Q_b,\quad
\{\Sigma,\tilde{Q}_a\}=\epsilon_{ab}\tilde{Q}_b.&\nonumber
\end{eqnarray}
For convenience of comparison with the fermion-monopole
system,
we do not include
a constant $\alpha$, playing a role of the central
element of the superalgebra, in the definition
of the $osp(2|2)$ spin integral $\Sigma$
generating the $so(2)\cong u(1)$ rotations
of the supercharges.
The quantity
\begin{equation}
C=4(D^2-KH)-2\alpha\Sigma+\alpha^2
\label{casim}
\end{equation}
is the Casimir element of the superalgebra
(\ref{osp22}) taking here the value $C=0$.

Since the system is described by the two independent
odd (Grassmann) variables $\psi_a$, $a=1,2$,
not all the odd
integrals (\ref{qs}) are independent.
Using the explicit form of the integrals
(\ref{hscm}), (\ref{tdk}), (\ref{qs}),
one finds that they satisfy the odd
and even relations
\begin{equation}
\alpha Q_a+2\epsilon_{ab}(Q_b D-\tilde{Q}_b H)=0,\qquad
\alpha \tilde{Q}_a+2\epsilon_{ab}(Q_bK-\tilde{Q}_bD)=0,
\label{qs1}
\end{equation}
\begin{equation}
Q_1Q_2-2i\Sigma H=0,\qquad
\tilde{Q}_1\tilde{Q}_2-2i\Sigma K=0,\qquad
Q_a\tilde{Q}_b+i\delta_{ab}\alpha \Sigma
-2i\epsilon_{ab}\Sigma D=0.
\label{qsas}
\end{equation}
The set of quadratic combinations defined by
Eqs. (\ref{qs1}), (\ref{qsas})
is  transformed linearly
under the action of the $osp(2,2)$ generators.
If the relations (\ref{qs1}) are treated as
a homogeneous linear set of equations for
$Q_a$ and $S_a$,
then due to the equality
$C=0$ and nilpotent character of $\Sigma$,
its
determinant
$\Delta=(C+2\alpha\Sigma)^2$
takes a zero value.
And vice versa, the condition
$\Delta=0$ for the determinant of the
homogeneous system of equations
for $Q_a$ and $\tilde{Q}_a$ (\ref{qs1})
fixes the value of the classical Casimir element: $C=0$.

It is useful to look at the symmetries
of the one-dimensional model
(\ref{hscm})
from the viewpoint of the planar system of
a free spin-$1/2$ particle
given by the action
\begin{equation}
A=\int\left (\frac{1}{2}\dot{x}_a^2-
\frac{i}{2}\dot\psi_a\psi_a\right)dt
\label{2d}
\end{equation}
and reduced to the surface of the fixed
full angular momentum \cite{AnPl}.
The system (\ref{2d})
can be characterized
by the set of quadratic integrals
\begin{equation}
H=\frac{1}{2}p_a^2,\quad
K=\frac{1}{2}X_a^2,\quad
D=\frac{1}{2}X_ap_a,\quad
\Sigma=-\frac{i}{2}\epsilon_{ab}\psi_a\psi_b,
\label{2dev}
\end{equation}
$$
L=\epsilon_{ab}X_ap_b,
$$
\begin{equation}
Q_1=p_a\psi_a,\quad
Q_2=\epsilon_{ab}p_a\psi_b,\quad
\tilde{Q}_1=X_a\psi_a,\quad
\tilde{Q}_2=\epsilon_{ab}X_a\psi_b
\label{2dod}
\end{equation}
constructed from the obvious set
of linear integrals of motion
$p_a$, $X_a=x_a-p_at$
and $\psi_a$.
The integrals (\ref{2dev}), (\ref{2dod})
form a superalgebra
of the form (\ref{osp22}),
in which instead of the parameter $\alpha$
the full angular momentum $J=L+\Sigma$
plays the role of the central element.
As a consequence,
the system
(\ref{hscm}) may be obtained by reducing
the system
(\ref{2d}) to the surface of the constraint
\begin{equation}
J-\alpha=0.
\label{cons}
\end{equation}
Due to zero Poisson brackets of
the quadratic integrals (\ref{2dev}),
(\ref{2dod})
with the constraint (\ref{cons}), they
are observables, and
after reduction their
superalgebra takes exactly the form (\ref{osp22}),
while the integrals themselves
take the form of the corresponding integrals
(\ref{hscm})--(\ref{qs})
(for the details see ref. \cite{AnPl}).

One can introduce into the system one more odd
degree of freedom described by the Grassmann variable
$\psi_3$ if to add the kinetic term
$-\frac{i}{2}\dot\psi_3\psi_3$ into
the Lagrangian of the system (\ref{2d}).
With such an extension, classically
the system will be described by one more odd integral
of motion $\Gamma=\psi_3$ being also,
due to the relation $\{\psi_3,J\}=0$,
the observable variable.
This additional odd integral
satisfies the relation $\{\Gamma,\Gamma\}=-i\cdot 1$,
and has zero Poisson brackets with all the
even and odd integrals (\ref{2dev}),
(\ref{2dod}).
With respect to
the superalgebra $osp(2|2)$,
the additional integral $\Gamma$ may be interpreted as a
classical analog of the
quantum grading operator.
Indeed, at the quantum level,
the Grassmann variables satisfying
the Poisson bracket relations
$\{\psi_a,\psi_b\}=-i\delta_{ab}$
are transformed into the
generators of the
Clifford algebra $Cl_2$
(for the initial system with $\psi_a$, $a=1,2$),
or of the $Cl_3$ (for the extended system with
the three odd variables $\psi_a$, $a=1,2,3$).
Here, in  correspondence with the relation
$dim\, Cl_{2n+1}=dim\, Cl_{2n}=2^{n}$
for the dimensions of irreducible representations,
in both cases the quantum analogs  of
the odd variables
can be realized in terms of Pauli matrices,
$\hat{\psi}_a=\sqrt{\frac{\hbar}{2}}\sigma_a$,
and $\sigma_3\propto \hat{\psi}_3$ is identified as
a grading operator of the quantum $osp(2|2)$ superalgebra.

Up to a numerical factor
the integral
$\hat{\Sigma}=-\frac{i}{2}[\hat{\psi}_1,\hat{\psi}_2]=
\frac{\hbar}{2}\sigma_3$ coincides with the
integral $\hat{\psi}_3$ of the extended system.
Hence, the integral
$\hat{\Sigma}=\frac{\hbar}{2}\sigma_3$
of the 1D superconformal model (\ref{hscm})
may be treated not only as an
even generator of the $osp(2|2)$,
but simultaneously it
may be considered as a grading operator
of the superconformal algebra.

Note that at the quantum level,
the condition of existence of a nontrivial
solution to the system of equations
(\ref{qs1})
is reduced to the equation
$$
\left(\hat{C}-\frac{3}{4}\hbar^2\right)
\left(\hat{C}-\frac{3}{4}\hbar^2+
8\hbar \alpha\sigma_3\right)=0,
$$
which fixes the value of the quantum analog of
the Casimir element (\ref{casim}),
\begin{equation}
\hat{C}\equiv 4\hat{D}^2-2(\hat{K}\hat{H}
+\hat{H}\hat{K})+\alpha(\alpha-\hbar\sigma_3)=
\frac{3}{4}\hbar^2.
\label{casqscm}
\end{equation}

\section{Dynamical symmetry of the fermion-monopole}

The 3D fermion-monopole system is described by the
Hamiltonian
\begin{equation}
H=\frac{1}{2}P_i^2-eB_iS_i
\label{hfm}
\end{equation}
with
$P_i=p_i-eA_i$,
$B_i=\epsilon_{ijk}\partial_jA_k=gx_i/|x|^3$,
$|x|=\sqrt{x_ix_i}$,
$S_j=-\frac{i}{2}\epsilon_{jkl}\psi_k\psi_l$,
and by the fundamental Poisson brackets
$\{x_i,p_j\}=\delta_{ij}$,
$\{\psi_j,\psi_k\}=-i\delta_{jk}$.
The Hamiltonian (\ref{hfm}) and the quantities
\begin{equation}
D=\frac{1}{2}X_iP_i+etB_iS_i=\frac{1}{2}x_iP_i-tH,
\label{dfm}
\end{equation}
\begin{equation}
K=\frac{1}{2}X_i^2-et^2B_iS_i=
\frac{1}{2}x_i^2-2tD-t^2H
\label{kfm}
\end{equation}
together with the full
angular momentum
$J_i$, given by the relations
\begin{equation}
J_i=L_i-\nu n_i+S_i,\qquad
L_i=\epsilon_{ijk}x_jP_k,\qquad
n_i=\frac{x_i}{|x|},\qquad
\nu=eg,
\label{jil}
\end{equation}
constitute the set of integrals
of motion generating the $so(1,2)\oplus so(3)$
symmetry \cite{Jackiw,mpmono}.
Here, the $so(1,2)$ Casimir element $D^2-KH$
and the $so(3)$ rotational invariant are related
by the equation
\begin{equation}
C=4(D^2-KH)+{\cal J}
-2{\cal Q} _L=0,
\label{3dcas}
\end{equation}
where
\begin{equation}
{\cal J}=J_i^2-\nu^2,
\label{jnu}
\end{equation}
\begin{equation}
{\cal Q} _L=L_iS_i.
\label{qls}
\end{equation}
The ${\cal Q} _L$ is the integral of motion
commuting with all the generators of the algebra
$so(1,2)\oplus so(3)$,
and extending it to the  $so(1,2)\oplus so(3)\oplus u(1)$.

The odd, $\psi_i$, and the even (the fermion's spin), $S_i$,
nilpotent vectors
satisfy the relations
\begin{equation}
\{S_i,\psi_j\}=\epsilon_{ijk}\psi_k,\qquad
\{S_i,S_j\}=\epsilon_{ijk}S_k,\qquad
\psi_iS_j=\frac{1}{3}\delta_{ij}(S_k\psi_k),
\label{xis}
\end{equation}
and their evolution is described by the same precession
motion,
\begin{equation}
\frac{d}{dt}\psi_i=e\epsilon_{ijk}\psi_j B_k,
\qquad
\frac{d}{dt}S_i=e\epsilon_{ijk}S_j B_k.
\label{press}
\end{equation}
Classically, they may be treated as ``parallel"
vectors, $\epsilon_{ijk}\psi_jS_k=0$,
and from eq. (\ref{press}) one finds that
their scalar product
\begin{equation}
{Q}_S=S_i\psi_i
\label{qss}
\end{equation}
is the odd  integral of motion.
Analogously to the 1D superconformal model,
in irreducible representation
the quantum operators $\hat{\psi}_i$
and $\hat{S}_i$ can be realized
in terms of the Pauli matrices,
$\hat{\psi}_i=\sqrt{\frac{\hbar}{2}}\sigma_i$,
$\hat{S}_i=\frac{\hbar}{2}\sigma_i$,
i.e., up to a numerical factor
they are represented by the same operators
$\sigma_i$.
Hence, the quantum analog of the
odd integral of motion (\ref{qss}) is reduced to a
pure quantum constant:
$\hat{Q}_S=3(\hbar/2)^{3/2}$.

Defining the vector
$X_i=x_i-tP_i$,
which in the free case
$eg=0$ is reduced to the
integral generating the Galilean boosts,
one can construct the pair of odd quantities
\begin{equation}
Q_P=P_i\psi_i,\qquad
Q_X=X_i\psi_i
\label{qpx}
\end{equation}
satisfying the  Poisson bracket relations
\begin{equation}
\{Q_P,Q_P\}=-2iH,\qquad
\{Q_X,Q_X\}=-2iK,\qquad
\{Q_X,Q_P\}=-2iD.
\label{khdfm}
\end{equation}
For the components of the vector $P_i$
the classical relation
\begin{equation}
\epsilon_{ijk}\{P_i,\{P_j,P_k\}\}\propto \delta^{(3)}(x)
\label{jacobi}
\end{equation}
takes place
(see also refs.
\cite{PesWei,KazYangGold,JackR}), and
the quantities (\ref{qpx}) are the odd integrals of motion
if the point $x_i=0$ is excluded from the
configuration space. In a physical context,
this can be achieved by some sort
of `regularization', e.g., by adding into the Hamiltonian
a spherical reflecting barrier potential $V(|x|)=+\infty$
for $|x|\leq a$,
and $V(|x|)=0$ for $|x|\geq a>0$, and by
taking subsequently a limit $a\rightarrow 0$; for
alternative
regularization see refs.
\cite{KazYangGold,Gold,Callias}.
Here we are not interested in
regularization and related problem
of bounded states
\cite{KazYangGold,Gold,Callias,Chandra,Lang},
and in what follows
will assume simply that
the point $x_i=0$ is excluded.

The odd integrals
$Q_P$ and $Q_X$,
being analogs of the integrals $Q_1$ and $\tilde{Q}_1$
of the superconformal model (cf.
the structure of the integrals (\ref{qpx})
with that of the integrals $Q_1$ and $\tilde{Q}_1$
given by eq. (\ref{2dod})),
generate together with the
$H$, $K$, $D$ and $J_i$
the $osp(1,2)\oplus so(3)$ superalgebra
with the Casimir element (\ref{3dcas}), which was
discussed by D'Hoker and Vinet \cite{DHV}.
In the fermion-monopole system,
however,
there is also the
nontrivial classical odd integral  (\ref{qss}).
Its bracket with the odd integrals (\ref{qpx})
generates new even integrals
\begin{equation}
{\cal Q} _P=P_iS_i,\qquad
{\cal Q} _X=X_iS_i.
\label{pxeven}
\end{equation}
The operators
corresponding to the classical quantities
(\ref{qpx}) coincide
up to a numerical factor with the quantum analogs of
(\ref{pxeven}). But classically the sets
(\ref{qpx}) and (\ref{pxeven}) are
different, and their mutual nontrivial
Poisson brackets
produce via the relation
$$
\{Q_X,{\cal Q} _P\}=\{{\cal Q} _X,Q_P\}=Q_L+Q_S
$$
a new integral of motion
\begin{equation}
Q_L=L_i\psi_i
\label{ql}
\end{equation}
being the odd  counterpart of the even
integral (\ref{qls}). Again, these two,
${\cal Q}_L$ and $Q_L$,
are related
by the Poisson bracket with the integral
(\ref{qss}),
$\{Q_S,Q_L\}=-3i{\cal Q} _L$,
and their quantum analogs are different only in a
numerical factor proportional to $\hbar^{1/2}$.

Let us change the
integral $Q_L$ for the linear combination of $Q_L$ and
$Q_S$,
\begin{equation}
Q_Y=Q_L+\frac{2}{3}\,Q_S.
\label{calqj}
\end{equation}
It is this odd integral,
satisfying the relation
$$
\{Q_Y,Q_Y\}=-i{\cal J},
$$
that was treated in Ref. \cite{MacHol}
as a generator of a new supersymmetry.
It, unlike the $Q_L$,
has zero Poisson bracket relations
(anticommutes at the quantum level)
with the odd integrals (\ref{qpx}).
On the other hand, the Poisson brackets
of the odd integrals (\ref{qpx}) with the
even integral ${\cal Q}_L$,
which due to the relation
$\{Q_S,Q_S\}=0$
may be treated as an even counterpart
of (\ref{calqj}) as well,
generate the two new odd integrals of motion,
$$
\{{\cal Q} _L,Q_P\}=Q_{\cal P},\qquad
\{{\cal Q} _L,Q_X\}=Q_{\cal X},
$$
where
\begin{equation}
Q_{\cal P}={\cal P}_i\psi_i,\qquad
Q_{\cal X}={\cal X}_i\psi_i,
\label{qpqx}
\end{equation}
\begin{equation}
{\cal P}_i= \epsilon_{ijk}L_jP_k+\frac{2}{3}\nu|x|^{-1}S_i,
\qquad
{\cal X}_i=\epsilon_{ijk}
L_jX_k-\frac{2}{3}t\nu|x|^{-1}S_i.
\label{calpx}
\end{equation}
The brackets of the integrals (\ref{qpqx})
with (\ref{qss})
produce their even counterparts
$$
{\cal Q} _{\cal P}={\cal P}_iS_i,\qquad
{\cal Q} _{\cal X}={\cal X}_iS_i.
$$
We list all the brackets of the
nilpotent integrals,
except the trivial brackets
$$
\{Q_S,Q_S\}=\{Q_S,{\cal Q}_L\}=0,
$$
in the table.

\begin{table}[h]
\begin{center}
\parbox{13cm}{Table. Poisson brackets between the nilpotent
integrals of motion. Here
$\Lambda \equiv
{\cal Q}_L+ {\cal J}$,
$Q_+\equiv Q_Y+\frac{1}{3}Q_S$. }
\vskip 2mm
\begin{tabular}{|ll|c|c|c|c|c|c|c|c|c|c|}
\hline
&&$Q_X$&$Q_P$&
$Q_{\cal X}$&$Q_{\cal P}$&$Q_Y$&
${\cal Q} _X$&${\cal Q}_P$&
${\cal Q} _{\cal X}$&
${\cal Q} _{\cal P}$
\\\hline\hline
$Q_X$
&&$-2iK$&$-2iD$&
$0$&$i\Lambda$&$0$&
$0$&
$Q_+$&
$2KQ_Y$&
$2DQ_Y$
\\\hline
$Q_P$
&&$-2iD$&$-2iH$&$-i\Lambda$&$0$&$0$&
$-Q_+$&
$0$&
$2DQ_Y$&
$2HQ_Y$
\\\hline
$Q_{\cal X}$
&&$0$&$-i\Lambda$&$-2i{\cal J}K$&
$-2i{\cal J}D$&$0$&
$-2KQ_Y$&
$-2DQ_Y$&
$0$&${\cal J}Q_+$
\\\hline
$Q_{\cal P}$
&&$i\Lambda$&$0$&$-2i{\cal J}D$&
$-2i{\cal J}H$&$0$&
$-2DQ_Y$&
$-2HQ_Y$&
$-{\cal J}Q_+$&$0$
\\\hline
$Q_Y$
&&$0$&$0$&$0$&$0$&$-i{\cal J}$&
$Q_{\cal X}$&
$Q_{\cal P}$&
$-{\cal J}Q_X$&$-{\cal J}Q_P$
\\\hline
${\cal Q}_L$
&&$Q_{\cal X}$&$Q_{\cal P}$&
$-{\cal J}Q_X$&$-{\cal J}Q_P$&$0$&
${\cal Q}_{\cal X}$&
${\cal Q}_{\cal P}$&$-{\cal J}{\cal Q}_X$&
$-{\cal J}{\cal Q}_P$
\\\hline
$Q_S$
&&$-3i{\cal Q}_X$&$-3i{\cal Q}_P$&
$-3i{\cal Q}_{\cal X}$&
$-3i{\cal Q}_{\cal P}$&$-3i{\cal Q}_L$&
$0$&$0$&$0$&$0$
\\\hline
${\cal Q} _X$
&&$0$&$Q_+$&
$2KQ_Y$&
$2DQ_Y$&-$Q_{\cal X}$&
$0$&
${\cal Q}_L$&$2K{\cal Q}_L$&
$2D{\cal Q}_L$
\\\hline
${\cal Q} _P$
&&$-Q_+$&$0$&
$2DQ_Y$&
$2HQ_Y$&$-Q_{\cal P}$&
$-{\cal Q}_L$&$0$&$2D{\cal Q}_L$&
$2H{\cal Q}_L$
\\\hline
${\cal Q} _{\cal X}$
&&$-2KQ_Y$&$-2DQ_Y$&
$0$&
$-{\cal J}Q_+$&${\cal J}Q_X$&
$-2K{\cal Q}_L$&
$-2D{\cal Q}_L$&$0$&
${\cal J}{\cal Q}_L$
\\\hline
${\cal Q} _{\cal P}$
&&$-2DQ_Y$&$-2HQ_Y$&
${\cal J}Q_+$&$0$&${\cal J}Q_P$&
$-2D{\cal Q}_L$&
$-2H{\cal Q}_L$&
$-{\cal J}{\cal Q}_L$&$0$
\\\hline

\end{tabular}
\end{center}
\end{table}

The quantities $Q_S$, $Q_Y$ and ${\cal Q} _L$
have zero Poisson brackets with the $so(1,2)$ generators
(\ref{hfm}), (\ref{dfm}), (\ref{kfm}),
while  the nontrivial  brackets of the odd integrals
(\ref{qpx}) with the $so(1,2)$ generators are
\begin{equation}
\{D,Q_P\}=\frac{1}{2}Q_P,\qquad
\{D,Q_X\}=-\frac{1}{2}Q_X,\quad
\{K,Q_P\}=Q_X,\qquad
\{H,Q_X\}=-Q_P.
\label{evod}
\end{equation}
The pairs
($Q_{\cal P}$, $Q_{\cal X}$),
(${\cal Q} _P$, ${\cal Q} _X$),
and (${\cal Q} _{\cal P}$,
${\cal Q} _{\cal X}$),
like the pair ($Q_P$, $Q_X$),
form the spin-1/2 representations
of the $so(1,2)$,
i.e. their corresponding brackets with
the $H$, $K$ and  $D$
have a form similar to (\ref{evod}).

The odd integrals
$Q_P$, $Q_X$, $Q_{\cal P}$ and $Q_{\cal X}$
satisfy the relations to be analogous
to the relations (\ref{qs1}), (\ref{qsas}):
\begin{eqnarray}
&Q_P{\cal J}+2(Q_{\cal P}D-Q_{\cal X}H)=0,\qquad
Q_X{\cal J}-2(Q_{\cal X}D-Q_{\cal P}K)=0,&\nonumber\\
&Q_{\cal P}-2(Q_PD-Q_XH)=0,\qquad
Q_{\cal X}+2(Q_XD-Q_PK)=0,&
\label{qqcal}
\end{eqnarray}
\begin{eqnarray}
&Q_PQ_{\cal P}-2iH{\cal Q} _L=0,\qquad
Q_XQ_{\cal X}-2iK{\cal Q} _L=0,\qquad
Q_PQ_X+2i{\cal Q} _L=0,&\nonumber\\
&Q_{\cal P}Q_{\cal X}-i{\cal Q} _L
{\cal J}=0,
\qquad
Q_PQ_{\cal X}+2i{\cal Q} _LD=0,
\qquad
Q_{\cal P}Q_X-2i{\cal Q} _LD=0.&
\label{qqhdk}
\end{eqnarray}
Having in mind this similarity, and
comparing
the structure of the Poisson brackets
of the integrals
$Q_P$, $Q_X$, $Q_{\cal P}$,
$Q_{\cal X}$,
$H$, $K$, $D$ and ${\cal Q}_L$
of the fermion-monopole system
with that for the integrals
$Q_1$, $\tilde{Q}_1$,
$Q_2$, $\tilde{Q}_2$,
$H$, $K$, $D$ and $\Sigma$
of the superconformal model,
we find that the former set
of the integrals forms
a nonlinear $\Z_2$-graded Poisson algebra
which can be considered as
a nonlinear generalization of the
superconformal algebra $osp(2,2)$.

Since ${\cal J}$ has zero Poisson brackets with
all the integrals, it plays the role of the
central charge of the nonlinear superconformal symmetry
to be analogous to that
for the $\alpha^2$ in the superconformal model.
From the comparison it follows also that the
odd integral $Q_Y$ is similar
to the odd integral $\Gamma$
playing the role of the classical analog
of the grading operator
for the superconformal model.
For $L_i^2\neq 0$,
the rescaling
\begin{equation}
Q_{\cal P}\rightarrow Q_{\cal P}{\cal J}^{-1/2},
\quad
Q_{\cal X}\rightarrow Q_{\cal X}{\cal J}^{-1/2},
\quad
{\cal Q}_L\rightarrow
{\cal Q}_L{\cal J}^{-1/2}
\label{scale}
\end{equation}
transforms the nonlinear superconformal
algebra of the fermion-monopole system
into the $osp(2,2)$
Lie superalgebra
(\ref{osp22}),
in which the role of the parameter $\alpha$
is played by the ${\cal J}^{1/2}$.

For $L_i^2\neq 0$,
one can define the vector
$$
N_i=
\left(
1-\frac{2\nu}{L_j^2}S_kn_k\right)Y_i,\qquad
Y_i=L_i+
\frac{2}{3}S_i.
$$
Then, using Eq. (\ref{jil}), the equality $S_iS_j=0$
and the last relation from
Eq. (\ref{xis}),
one can represent the integrals
$Q_{\cal P}$,
$Q_{\cal X}$,
$Q_Y$ and ${\cal Q}_L$
in the form
\begin{equation}
Q_{\cal P}=\epsilon_{ijk}N_iP_j\psi_k,\quad
Q_{\cal X}=\epsilon_{ijk}N_iX_j\psi_k,\quad
{\cal Q}_L=-\frac{i}{2}\epsilon_{ijk}N_i\psi_j\psi_k,\quad
Q_Y=N_i\psi_i.
\label{qnpsi}
\end{equation}
The `extended' superconformal model
may also be represented in a 3D form by
introducing the notations
$x_i=(x_1,x_2,0)$,
$p_i=(p_1,p_2,0)$,
$\psi_i=(\psi_1,\psi_2,\psi_3)$
and  $N_i=(0,0,1)$.
Then the integrals (\ref{2dev}),
(\ref{2dod})
can be rewritten in the form of
3D scalar products,
similar to the integrals of motion
of the fermion-monopole system.
In particular, the classical analog
of the grading operator,
$\Gamma=\psi_3$,
takes a form similar to $Q_Y$
from (\ref{qnpsi}),
$\Gamma=N_i\psi_i$.

Identifying the $\Z_2$-graded Poisson algebra
of the integrals of motion as a nonlinear superconformal
algebra (plus the `decoupled' $so(3)\cong su(2)$),
we have omitted from consideration
the even nilpotent integrals
${\cal Q}_X$, ${\cal Q}_P$,
${\cal Q}_{\cal X}$ and ${\cal Q}_{\cal P}$.
This was done having in mind the quantum case,
where these integrals are different
from the quantum analogs of the corresponding
odd integrals only in a simple numerical
factor $\sqrt{{\hbar}/{2}}$.

Let us note here that the even nilpotent integrals
for the model of superconformal mechanics
extended by the odd integral $\psi_3$
might also  be constructed in an obvious way.
Having in mind the correspondence
$\alpha^2\sim {\cal J}$,
$\Sigma\Gamma \sim \frac{1}{3}Q_S$,
$\alpha\Gamma\sim Q_Y$,
$\alpha\Sigma\sim {\cal Q}_L$
and relations (\ref{scale}),
one finds
that the even nilpotent quantities
$i\Gamma Q_2$, $-i\alpha\Gamma Q_1$,
$i\Gamma \tilde{Q}_2$ and
$-i\alpha\Gamma \tilde{Q}_1$ would be the analogs
of the integrals
${\cal Q}_P$, ${\cal Q}_{\cal P}$,
${\cal Q}_X$ and ${\cal Q}_{\cal X}$,
respectively, which,
with taking into account Eq. (\ref{qsas}),
would
generate the nonlinear Poisson bracket relations
of the form presented in the Table.
One could treat
the complete set of these classical `commutation' relations
of the all even and odd integrals
as some nonlinear Poisson superalgebra ${\cal G}$
 \cite{nonlin}.
However, its nature is essentially  different from
that of the nonlinear generalization
of the $osp(2|2)\oplus su(2)$ generated by the
integrals $H$, $K$, $D$, ${\cal Q}_L$,
$Q_P$, $Q_X$, $Q_{\cal P}$, $Q_{\cal X}$ and $J_i$
(plus the Yano supercharge $Q_Y$ \cite{MacHol,GRH}
being the classical
analog of the grading operator of the
quantum version of this superalgebra, see below).
The difference is that the nonlinearity
of the generalized $osp(2|2)\oplus su(2)$
is encoded in the central charge
${\cal J}$ appearing additively and multiplicatively
in the Poisson bracket relations,
while
the nontrivial $so(1,2)$ generators $H$, $D$
and $K$ appear as multiplicative factors in the
Poisson superalgebra ${\cal G}$.
One of the consequences of this is
that there exists no analog of
the rescaling procedure (\ref{scale}) which
would transform ${\cal G}$  into some Lie superalgebra.

The role of the quantum $osp(2|2)$ spin
operator is played by
\begin{equation}
\hat{\cal Q}_L=\frac{\hbar}{2}\left(\hat{L}_i\sigma_i+
\hbar\right).
\label{opql}
\end{equation}
In correspondence with the classical Poisson bracket
relations,
it rotates the supercharges (cf. with the two last relations
from (\ref{osp22})),
\begin{eqnarray}
&[\hat{\cal Q}_L,\hat{Q}_X]
=i\hbar\hat{Q}_{\cal X},
\quad
[\hat{\cal Q}_L,\hat{Q}_P]
=i\hbar\hat{Q}_{\cal P},&
\nonumber\\
&[\hat{\cal Q}_L,\hat{Q}_{\cal X}]
=-i\hbar\hat{\cal J}\hat{Q}_X,
\quad
[\hat{\cal Q}_L,\hat{Q}_{\cal P}]
=-i\hbar\hat{\cal J}\hat{Q}_P.&
\label{sigmaq}
\end{eqnarray}
Here
$$
\hat{\cal J}=\hat{J}_i^2-
\nu^2+\frac{1}{4}\hbar^2
$$
is the quantum analog of (\ref{jnu}) with the quantized
$\nu$, $|\nu|=\hbar n/2$,
$n\in \N$,
and the quantum analogs of $Q_{\cal X}$ and
$Q_{\cal P}$ are obtained from (\ref{qpqx}),
(\ref{calpx})
by a simple antisymmetrization
of  the noncommuting
factors $\hat{L}_j$ and $\hat{P}_k$, or $\hat{L}_j$
and $\hat{X}_k$.

The only difference  of the nonlinear
superalgebra formed by the fermion-monopole  integrals of
motion  in comparison with  the quantum version of the
$osp(2|2)$ superalgebra (\ref{osp22}) consists in the
presence
of the additional factor $\hat{\cal J}$
in commutators (\ref{sigmaq})
and in anticommutators
of $\hat{Q}_{\cal P}$ and  $\hat{Q}_{\cal X}$
(see the table).
In accordance with the classical Poisson
bracket relations we have also
$$
[\hat{Q}_P,\hat{Q}_{\cal X}]_{{}_+}=
[\hat{Q}_{\cal P},\hat{Q}_X]_{{}_+}=
\hbar\,(\hat{\cal Q}_L+\hat{\cal J}).
$$
The integral
$
\hat{\cal Q}_L
$
commutes with all the even generators of
the nonlinear generalization of the
$osp(2|2)\oplus su(2)$,
and in addition,  due to the relation
\begin{equation}
\hat{\cal Q}_L
=\sqrt{\frac{\hbar}{2}}
\hat{Q}_Y
\label{qly}
\end{equation}
anticommutes with the odd supercharges.
Satisfying the relation
\begin{equation}
\hat{\cal Q}_L^2=\frac{\hbar^2}{4}\hat{\cal J},
\label{calq2}
\end{equation}
it is the square root from the
central element $\hat{\cal J}$
of the nonlinear superalgebra,
and may be treated simultaneously
as the grading operator
of the nonlinear version of
the $osp(2|2)\oplus su(2)$.

The quantum analog of  the classical
relation (\ref{3dcas}) is
$$
\hat{C}=4\hat{D}^2-2(\hat{K}\hat{H}+\hat{H}\hat{K})
+\hat{\cal J}-2\hat{\cal Q}_L=
\frac{3}{4}\hbar^2.
$$
This quantum relation  fixing the value of
$\hat{C}$
(cf. with  (\ref{casqscm}))
appears
as the condition of existence of nontrivial solutions
to the system of homogeneous equations
being the quantum analog of (\ref{qqhdk}).

In representation where the
squared full angular momentum operator
is diagonal,
$\hat{J}_i^2=j(j+1)\hbar^2$,
$j+\frac 12=\hbar^{-1}|\nu|+m$,
$m=0,1,2,...$,
we have $\hat{\cal J}=(|\nu|+m)^2-\nu^2$
\cite{DHV,mpmono}.
Then,
in accordance with relation (\ref{calq2}),
for the states with $m>0$
the appropriately normalized
operators  $\hat{\cal Q}_L$,
$\hat{Q}_{\cal X}$ and $\hat{Q}_{\cal P}$
(see eq. (\ref{scale})) together with the
rest of  integrals of motion generate the
Lie superalgebra $osp(2|2)\oplus su(2)$.
However, as in the case of the Kepler problem,
such a normalization procedure
has a hidden nonlocal nature.
Following  ref. \cite{DHV}, one can show
that in the sector  $m=0$, corresponding
classically to the phase space surface
given by the equations $L_i=0$, $S_jn_j=0$,
the symmetry of the system
is reduced to the
conformal symmetry $so(1,2)$.

Note that the analogy
with the Kepler problem,
mentioned in the Introduction,
is given by the quantum relations
(\ref{calq2}),
\begin{equation}
[\hat{Q}_P,\hat{Q}_P]_{{}_+}=2\hbar \hat{H},\quad
[\hat{\cal Q}_L,\hat{Q}_P]=i\hbar
\hat{Q}_{\cal P},
\quad
[\hat{Q}_{\cal P},\hat{Q}_{\cal P}]_{{}_+}=
2\hbar \hat{\cal J}\hat{H},
\label{Kep}
\end{equation}
see Eq. (\ref{sigmaq}) and the Table.
Here, however, it is necessary to
stress that
the relation (\ref{calq2}) is satisfied by the
$\hat{\cal Q}_L$ not as by
the even generator
of the nonlinear superconformal algebra,
but as its grading operator which
coincides (up to the quantum
factor, see Eq. (\ref{qly}))
with the Yano supercharge $\hat{Q}_Y$
\cite{MacHol,GRH}.

\section{Discussion and outlook}

To conclude, let us summarize the obtained results
and discuss shortly some problems that deserve a further
attention.

Comparing the system of the
charged fermion in the field of the
Dirac magnetic monopole with the model of
superconformal mechanics, we have showed that
its integrals of motion generate a nonlinear
$\Z_2$-graded  Poisson algebra, or quantum super- algebra,
which may be treated as a nonlinear generalization of the
$osp(2|2)\oplus su(2)$. In this nonlinear superalgebra,
containing the  $osp(1|2)$ \cite{DHV} as a Lie
sub-superalgebra, the shifted square of the full angular
momentum of the system plays the role of the central charge
appearing additively and multiplicatively in the
quantum (anti)commutation and classical Poisson bracket
relations.
The square root from the central charge is the
$osp(2|2)$ spin generating the $u(1)$ rotations of the odd
supercharges. Classically, it has an odd counterpart, whose
quantum analog is, up to a numerical factor
$\sqrt{{\hbar}/{2}}$,  the same $osp(2|2)$ spin
operator. As an odd integral, it generates a nonlinear
supersymmetry discussed earlier in
\cite{MacHol,Spec,PlFM}.
Since it anticommutes with all other odd
supercharges, and commutes with all the even integrals
 of the nonlinearly generalized
$osp(2|2)\oplus su(2)$ (including itself), it may be
identified as
a grading operator of the superalgebra.
Note also that the form of the nonlinear superalgebra
can be simplified a little bit by
shifting the $osp(2|2)$ spin
for the central charge,
${\cal Q}_L\rightarrow {\cal Q}_L+{\cal J}\equiv \Lambda$
(see the table).

The natural question is what are the possible physical
consequences of the described nonlinear superconformal
symmetry of the charge-monopole system?
Having in mind the mentioned analogy
with the hidden
symmetry of the Kepler problem,
it would be interesting to look
at the charge-monopole scattering problem \cite{KazYangGold}
from the perspective of the revealed symmetry.

In  refs. \cite{LP,AnPl} it was observed that
the change of a boson-fermion coupling constant
$\alpha\rightarrow n\alpha$,
$n\in\N$,
in the superconformal mechanics model corresponds
to the change of the particle's spin $\hbar/2$ for
$n\hbar /2$, and that the modified superconformal model is
characterized by the nonlinear superconformal symmetry
$osp(2|2)_n$, in which the set of $2(n+1)$  odd integrals
constitute the spin-$\frac{n}{2}$ representation of
the $so(1,2)$ \cite{LP,AnPl}.
Proceeding from the  close similarity between the
fermion-monopole system and superconformal mechanics model,
one could expect the appearance of
some generalization of the nonlinear superconformal
symmetry $osp(2|2)_n$ of refs. \cite{LP,AnPl}
as a symmetry for a higher spin charged particle in the
field of the Dirac
monopole.

\vskip 0.5cm
{\bf Acknowledgements}
\vskip 5mm
The work has been supported in part
by CONICYT-Chile (CL), and by
FONDECYT-Chile (grant 1010073) and by DICYT-USACH (MP).

\end{document}